\documentclass[superscriptaddress,letterpaper,twoside,twocolumn,showpacs,amsmath,amssymb,aps,prl,floatfix,showkeys,reprint,]{revtex4-1}
\usepackage{graphicx}
\usepackage{dcolumn}
\usepackage{bm}
\usepackage{amsmath}
\usepackage{amssymb}
\usepackage{multirow}
\usepackage{color}
\usepackage{textcmds} 
\usepackage{booktabs}
\usepackage{tabularx}
\usepackage[english]{babel}
\usepackage[utf8]{inputenc}

\usepackage[normalem]{ulem}
\usepackage{xcolor}

\begin{document}
\title{Domain Walls in a Row-Wise Antiferromagnetic Mn Monolayer}

\author{Jonas Spethmann}
\affiliation{Department of Physics, University of Hamburg, 20355 Hamburg, Germany}
\author{Martin Gr\"unebohm}
\affiliation{Department of Physics, University of Hamburg, 20355 Hamburg, Germany}
\author{Roland Wiesendanger}
\affiliation{Department of Physics, University of Hamburg, 20355 Hamburg, Germany}
\author{Kirsten von Bergmann}
\affiliation{Department of Physics, University of Hamburg, 20355 Hamburg, Germany}
\author{Andr\'e Kubetzka}
\email[Corresp.\ author: ]{kubetzka@physnet.uni-hamburg.de}
\affiliation{Department of Physics, University of Hamburg, 20355 Hamburg, Germany}
	   
\date{\today}

\begin{abstract} 
We investigate magnetic domain walls in a single fcc Mn layer on Re(0001) employing spin-polarized STM, atom manipulation, and spin dynamics simulations. The low symmetry of the row-wise antiferromagnetic (1$Q$) state leads to a new type of domain wall which connects rotational 1$Q$ domains by a transient 2$Q$ state with characteristic $90^\circ$\,angles between neighboring magnetic moments. The domain wall properties depend on their orientation and their width of about 2\,nm essentially results from a balance of Heisenberg and higher-order exchange interactions. Atom mani\-pulation allows domain wall imaging with atomic spin-resolution, as well as domain wall positioning, and we demonstrate that the force to move an atom is anisotropic on the 1$Q$ domain.
\end{abstract}
\maketitle

Antiferromagnets (AFMs) as first conceived by Louis N\'eel consist of two identical, interpenetrating ferromagnetic sublattices, magnetized in opposite directions; they are \qq{extremely interesting from the theoretical standpoint but do not appear to have any practical applications}~\cite{neelScience1971}. Nowadays, a large variety of magnetic systems with vanishing net magnetization falls into this category, including \qq{synthetic} AFMs~\cite{duineNP2018,finco2020}, collinear~\cite{kubetzkaPRL2005,kronleinPRL2018}, non-collinear~\cite{bodeN2007,gaoPRL2008b,nakatsujiN2015}, and non-coplanar systems~\cite{kurzPRL2001,spethmannPRL2020}, and AFMs are envisioned to play a dominant role in future spintronic devices~\cite{jungwirthNN2016,baltzRMP2018,lothS2012}.   
AFMs are versatile materials, which interact and coexist with superconductors~\cite{bobkovaPRL2005, knebelCRP2011}, display different and generally faster dynamics than ferromagnets (FM)~\cite{fiebigJPD2008}, and show distinct transport properties which can depend on their topology~\cite{smejkalNP2018}. Though a number of experimental techniques have been developed and refined to image AFM domains and domain wall (DW) positions, determining DW widths or resolving the spin configuration within a DW\,\cite{bodeNM2006} remains a challenging task~\cite{cheongNPJ2020}. Theoretical investigations of AFM DWs have focused on dynamic properties \cite{gomonayNP2018}, such as Lorentz contraction, suppression of Walker breakdown~\cite{shiinoPRL2016,gomonayPRL2016} or interactions with spin waves~\cite{lanNC2017,qaiumzadehPRB2018}.
The considered DWs, e.g.\ phase DWs which connect translational domains \cite{cheongNPJ2020}, are typically described by a coherent rotation of all magnetic sublattices. Whereas the \textit{static} profiles of these DWs are identical to DWs in equivalent FM systems~\cite{papanicolaouPRB1995,ulloaPRB2016}, it has been pointed out that the large variety of AFM states should allow for a wider range of AFM DW configurations compared to ferromagnets~\cite{baltzRMP2018}.

New AFM DW types, void of an FM counterpart, can emerge in systems where the symmetry of the AFM spin texture is lower than the supporting crystal lattice~\cite{hagemeisterPRL2016,kronleinPRL2018}, because the resulting rotational domains of the \textit{spin texture} have no FM analog. Here, we investigate DWs in the fcc-stacked hexagonal manganese (Mn) layer on Re(0001), which hosts a row-wise AFM ($1Q$) state, that was characterized in Ref.\,\onlinecite{spethmannPRL2020} with density functional theory (DFT) and spin-polarized (SP)-STM~\cite{bodeRPP2003,wiesendangerRMP2009}: the $1Q$ state is the result of an antiferromagnetic nearest neighbor Heisenberg exchange coupling, $-J_1(\mathbf{S}_i \cdot \mathbf{S}_j)$ with $J_1<0$, exchange frustration fulfilling $1>J_2/J_1>1/8$~\cite{hardratPRB2009}, where $J_2<0$ denotes the coupling strength to the next-nearest neighbor, and a small net contribution from higher order exchange interactions (HOIs)~\cite{hoffmannPRB2020}, which lowers the energy of the $1Q$ state with respect to the otherwise degenerate $2Q$ and $3Q$ state~\cite{kurzPRL2001,spethmannPRL2020}. In addition, the system has an easy-plane crystal anisotropy, $K<0$, and both the magnetic dipolar interaction and the anisotropic symmetric exchange (ASE) couple the spin direction to the spin texture, i.e.\ the AFM atomic rows, see schematics in Fig.\,1(a).

The DWs occur between rotational domains of the spin texture and show characteristic hexagonal patterns in SP-STM images. They can be modeled based on DFT parameters assuming a small positive effective HOI value, which determines the wall widths. The simulations show that in contrast to traditional AFM DWs, specific spin pairs rotate in \textit{opposite} directions which leads to a transient 2$Q$ state with characteristic 90$^\circ$ angles in the wall center. Using a Co adatom as a local sensor of the spin texture, the transient 2$Q$ DW structure is imaged with atomic spin-resolution and we finally demonstrate as a proof of principle, that AFM DW positions can be controlled by manipulating individual atoms.

\begin{figure}[htb] 
	\centering
	\includegraphics[width=0.99\columnwidth]{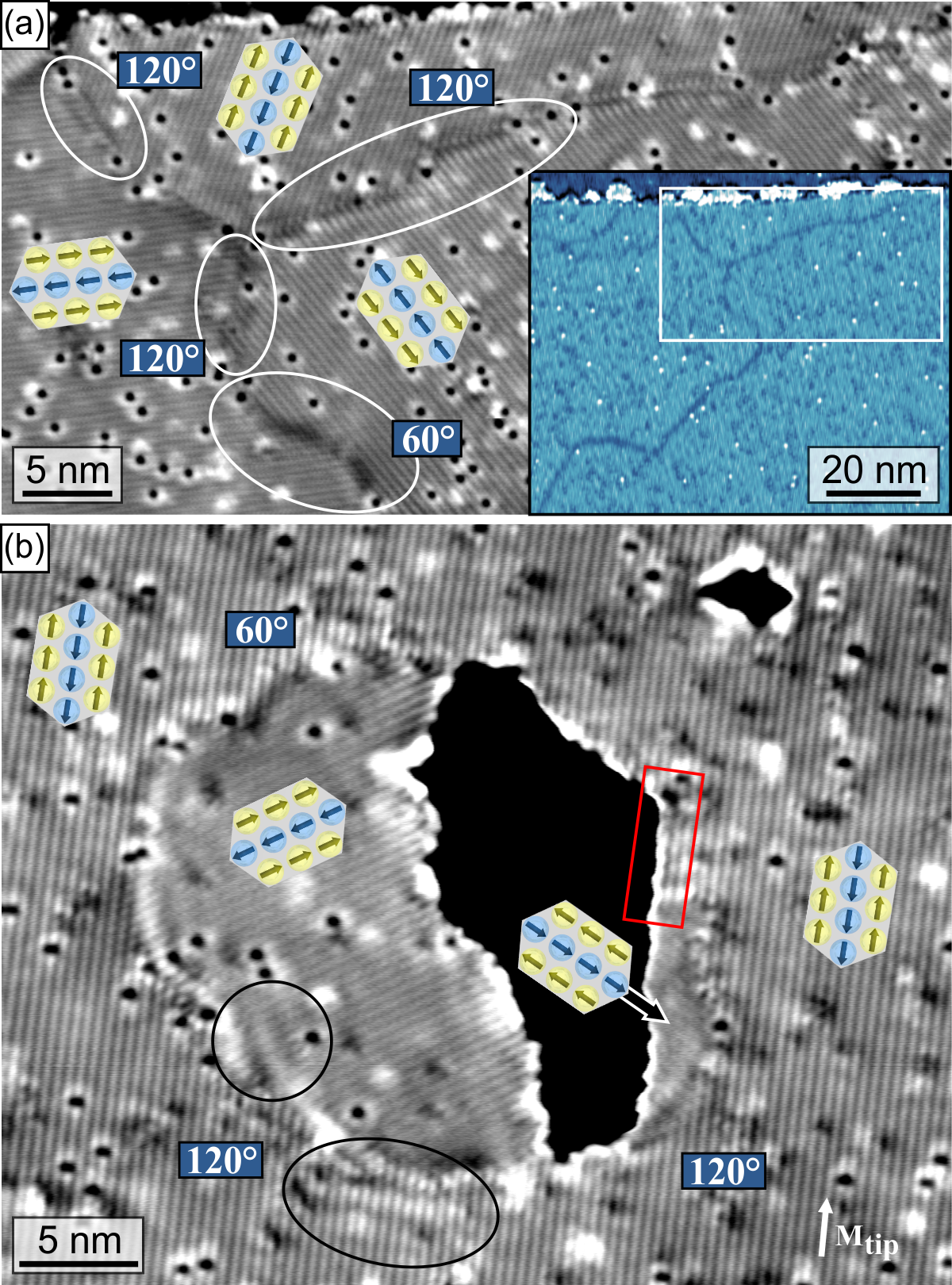}
	\caption{Magnetic domain configuration and electronic DW contrast. (a)~Constant current SP-STM image (height data) of three adjacent 1$Q$ rotational domains ($U=-30$\,mV, $I=7$\,nA). The DW directions vary, affected by atomic defects. The inset shows a larger field of view ($dI/dU$ map, $U=+500$\,mV, $I=3$\,nA) where the DWs are imaged without resolving the AFM spin texture. (b)~A vacancy island (black) with two additional domains which avoid AFM rows running along the edges; an exception is marked with a red frame ($U=+50$\,mV, $I=7$\,nA). Specific wall orientations lead to interference patterns, see black circles.}
	\label{fig1}
\end{figure}

We use a home-built STM at $T=4.2$\,K, equipped with a Cr bulk tip, which---depending on its \textit{in-situ} treatment---can assume an arbitrary magnetization direction with varying degrees of spin-polarization. In addition to the Mn/Re(0001) sample preparation described in Ref.\,\onlinecite{spethmannPRL2020}, for manipulation experiments we deposit single Co or Ir atoms onto the cold sample surface.
Figure 1(a) shows a surface area with native defects and three rotational magnetic domains, where the AFM atomic rows are resolved by spin-polarized tunneling. For the majority of DWs---in Fig.\,1 and in general---the AFM rows of adjacent domains enclose an angle of 120$^\circ$ at the DW position, whereas 60$^\circ$ is less common, see labels in Fig.\,1(a), referred to as 120$^\circ$ and 60$^\circ$\,DWs in the following. Individual DW orientations vary and apparently are influenced by atomic defects. Phase DWs, as found in Fe/W(001)~\cite{bodeNM2006}, which connect translational AFM domains by a 180$^\circ$ \textit{spin} rotation, are rare in fcc Mn/Re(0001), see Fig.\,S1 of the Supplemental Material~\cite{suppl}. At a larger scale, see inset of Fig.\,1(a), DWs can be imaged without atomic scale resolution by measuring differential conductance, indicating a sizeable influence of the DWs onto the spin-averaged local density of states due to their distinct spin texture \cite{hannekenNN2015}, see also Fig.\,S2 \cite{suppl}.
Figure 1(b) shows a typical example where additional domains exist at the sample boundary, in this case induced by a vacancy island (black). 
It seems that rotational domains with the AFM rows running along the edge, i.e.\ FM edges, are avoided, with an exception marked by a red frame.
We show in an idealized scenario, see Fig.\,S3 \cite{suppl}, that the effect results from $J_1/J_2>1$, i.e.\ from the asymmetry of the exchange interactions. It can be viewed as an AFM analog to domain formation by magnetic charge avoidance in FM systems and is a direct consequence of the low symmetry of the 1$Q$ state.

\begin{figure}[tbh] 
	\centering
	\includegraphics[width=0.99\columnwidth]{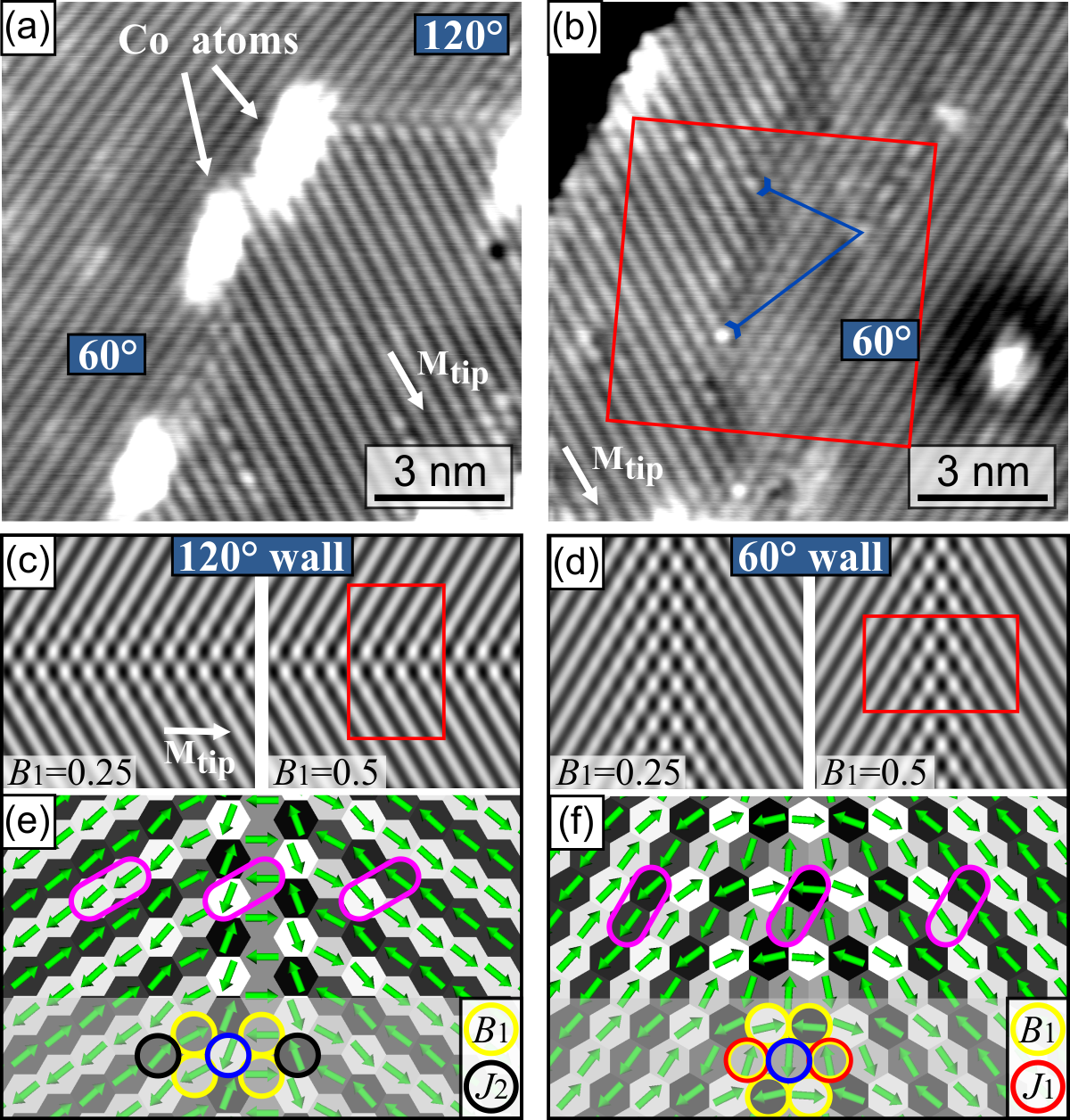}
	\caption{Domain wall types. (a)~SP-STM image of a DW with Co atoms or clusters with a 60$^\circ$ and a 120$^\circ$\,DW section and (b) a neighboring 60$^\circ$\,DW about $50$\,nm away imaged with the same tip ($U=+14.8$\,mV, $I=5$\,nA). Red frame indicates the area shown in Fig.\,3(a). (c)~$6\times 6$\,nm$^2$ SP-STM images calculated from a simulated 120$^\circ$\,DW and (d) 60$^\circ$\,DW, for two different values of the biquadratic term $B_1$. (e),(f) Atomic spin configuration within the red rectangles in (c), (d), respectively. Simulations are based on simplified DFT parameters (in meV/atom): $J_1=-25$, $J_2=-5$, $J_{\rm ASE}=+0.025$, and $K=-1$. For an arbitrary site (blue circle), interactions contributing to the DWs are marked with circles. Spin pairs rotate in \textit{opposite} directions across both walls, see pink frames~\cite{suppl}.}
	\label{fig2}
\end{figure}

\begin{table}[tb] 
\centering
\caption {AFM domain wall widths and energies in small angle approximation and for $B_1\ll J_1$~\cite{suppl}. We use a minimal model with only $J_1$, $J_2$, and $B_1$ (all in meV/atom) on a hexagonal layer with lattice constant $a$.
180$^\circ$ FM DWs are shown for comparison, with easy-axis anisotropy, $K>0$, in meV/atom and exchange stiffness $\textsc{a}$ and anisotropy $\textsc{k}$ in SI units.}
\begin{ruledtabular}
\begin{tabular}{l|l|l|l} 
DW Type & Model & Width & Energy\\
\hline
180$^\circ$ FM & $\textsc{a}$, $\textsc{k}$ (continuum) & $2\sqrt{\textsc{a}/ \textsc{k}}$ & $4\sqrt{\textsc{ak}}$\\
180$^\circ$ FM & $J_1>0$, $K>0$ & $2a\sqrt{\frac{3}{2} J_1/K} $ & $\frac{4}{a}\sqrt{2 J_1 K}$\\
120$^\circ$ AFM & $J_1,J_2<0$, $B_1>0$ & $\frac{a}{2}\sqrt{3|J_2|/B_1}$ & $\frac{8}{a}\sqrt{|J_2| B_1}$\\
60$^\circ$  AFM & $J_1,J_2<0$, $B_1>0$ & $\frac{a}{2}\sqrt{|J_1|/B_1}$ & $\frac{8}{a}\sqrt{\frac{1}{3} |J_1| B_1}$\\
\end{tabular}
\end{ruledtabular}
\label{table}
\end{table}

\begin{figure*}[tbh] 
	\centering
	\includegraphics[width=0.9\textwidth]{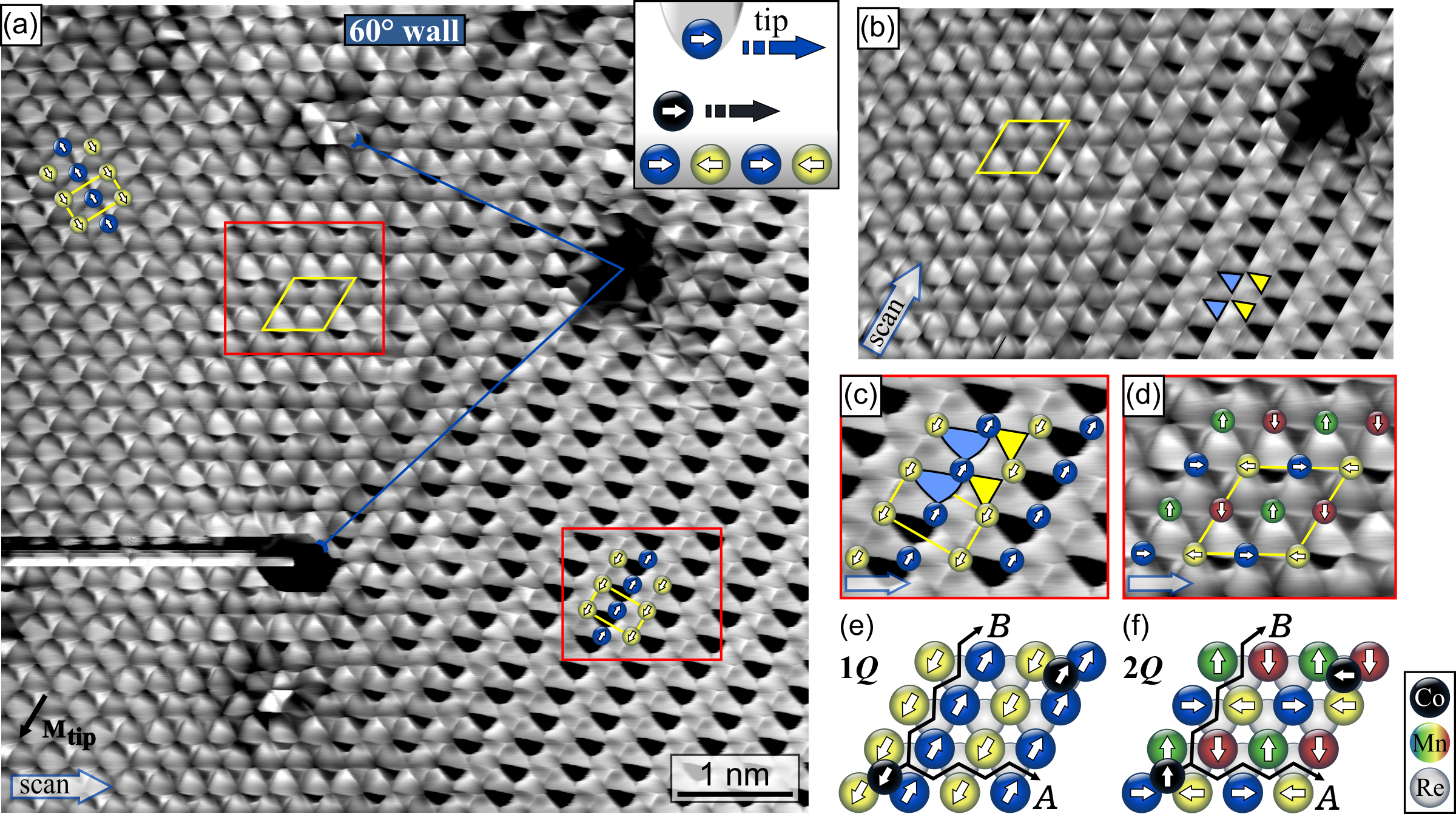}
	\caption{Transient 2$Q$ DW structure. (a)~Atom manipulation image (height data) of the area indicated in Fig.\,2(b), using a Co atom ($U=+3.8$\,mV, $I=60$\,nA), see inset schematics. Magnetic unit cells are indicated by yellow rectangles and rhomboid. Blue triangle marks three defects to assist comparison with Fig.\,2(b). Two 1$Q$ rotational domains and the transient 2$Q$ state are resolved. (b)~Changing the scanning direction by 60$^\circ$ has little influence on the measured data, except that the marked plaquettes, blue and yellow, have similar sizes. (c)~1$Q$ domain area from (a) with proposed Mn atom positions. Marked plaquettes have distinct sizes. (d)~DW region from (a) with $p(2\times 2)$ magnetic unit cell and proposed Mn atom spin texture. (e),(f) Schematic views of 1$Q$ and 2$Q$ state with proposed Re atom positions and inequivalent manipulation paths $A$ and $B$.}
	\label{fig3}
\end{figure*}

Figure 2(a) shows a closer view of a DW decorated with Co atoms, consisting of a 120$^\circ$ and a 60$^\circ$\,DW section. Both DWs show a narrow hexagonal pattern, which is more pronounced for the 60$^\circ$\,DW. A second 60$^\circ$\,DW in Fig.\,2(b), imaged under identical conditions, shows a more extended hexagonal transition region, probably as a result of a different local configuration of native defects. These experimental DWs are well reproduced by atomistic spin dynamics simulations~\cite{github} based on a simplified set of DFT parameters~\cite{spethmannPRL2020}, when a small biquadratic term, $-B_1(\mathbf{S}_i \cdot \mathbf{S}_j)^2$ , with $B_1=+0.5\pm 0.3$ meV/atom is chosen, representing the net contribution from the fourth-order HOIs. See also Fig.\,S4~\cite{suppl} for more details of the simulations. The respective calculated SP-STM images~\cite{heinzeAPA2006} are shown in Fig.\,2(c),(d) for two different values of $B_1$.
The simulated spin configurations in Fig.\,2(e),(f) show that the DWs accomplish two things: (i) the spin quantization axis rotates as in any FM DW and in addition (ii) the spin texture, i.e.\ the AFM atomic rows change direction. The latter is achieved by an \textit{opposite} rotation of adjacent spin pairs across the walls, see pink frames in Figs.\,2(e),(f). This leads to a transient 2$Q$ state in the center, which in its extended form consists of close-packed AFM rows with an inter-row angle of 90$^\circ$, see Ref.\,\onlinecite{KurzPHD} and Fig.\,3(f). Due to the AFM periodicity along the DW, both wall types have a vanishing net magnetization, and a number of pair-wise interaction terms does not contribute to the DW energy and shape: for a spin at an arbitrary site, marked by a blue circle in Fig.\,2(e),(f), we have indicated the partner sites actually contributing to the energy difference between domain and DW. Thus, for symmetry reasons 120$^\circ$\,DWs do not depend on $J_1$ and 60$^\circ$\,DWs do not depend on $J_2$.
Furthermore, it turns out that the three fourth-order HOI terms \cite{hoffmannPRB2020}, two-site ($B_1$) \cite{brownPRB1971}, three-site ($Y_1$) \cite{kronleinPRL2018}, and four-site ($K_1$) \cite{heinzeNP2011} four spin interaction, lead to almost identical DW profiles in our simulations~\cite{github}, because all terms scale with $\cos^2(2\alpha)$ in a 1$Q$--2$Q$ transformation, see Ref.\,\onlinecite{meyerPRB2017} and Fig.\,S5~\cite{suppl}. The $\cos^2(2\alpha)$ scaling leads to DW profiles closely following $\tanh$ functions, see Figs.\,S6--8~\cite{suppl}, for both DW types and allows a mapping onto analytical formulas for DW width and energy, shown in Table\,I. Here, the HOI term has a similar effect as the crystal anisotropy in an FM DW, but as a spin--spin interaction it does not provide coupling to the crystal lattice \cite{suppl}.
For fcc Mn/Re(0001) with $J_1/J_2\approx 4.5$ these formulas show that 60$^\circ$\,DWs are higher in energy and wider compared to 120$^\circ$\,DWs, both by a factor of $\sqrt{4.5/3}\approx 1.2$, in agreement with the larger overall length and smaller widths of 120$^\circ$\,DWs in our experimental data.

It is challenging to precisely determine the intrinsic DW width and shape from SP-STM measurements because the apparent DW profiles depend on the DW orientation and seem to be affected by defects, leading to variations in the regime of 1.5--2.5\,nm for 60$^\circ$\,DWs, see Fig.\,2(a),(b). Furthermore, as seen in Fig.\,1 and Fig.\,S2 \cite{suppl}, additional LDOS variations which occur in a wide bias regime complicate extracting the exact spin configuration. Hence, based on our SP-STM data we cannot exclude small perpendicular spin components in the DW, and can therefore not rule out a distortion of the coplanar 2$Q$ toward a 3$Q$ state, e.g.\ as a result of HOI terms beyond fourth-order \cite{brinkerPRR2020}.

To explore the Mn layer from a different angle we employ magnetic atom manipulation imaging~\cite{wolterPRL2012,ouaziPRL2014}, where a magnetic atom on the surface, following the moving tip while jumping from one lattice site to the next, is used as a local sensor and amplifier. The resulting data in Fig.\,3(a), measured across the 60$^\circ$ DW as indicated in Fig.\,2(b), is more complex than standard SP-STM, because the manipulated Co atom introduces additional degrees of freedom which contribute non-linearly to the tunnel current. On non-magnetic hexagonal surfaces the type of hollow site (fcc or hcp) in which the adatom is residing can be discriminated by a difference of apparent height, size and symmetry of the corresponding plaquettes in the manipulation images~\cite{stroscioS2004,ouaziSS2014}. This is also the case here and best seen for the left 1$Q$ domain in Fig.\,3(a) or the DW area in Fig.\,3(d): the larger upward pointing triangular plaquettes correspond to the energetically favored hollow sites, the smaller downward pointing triangles indicate the less favored sites, see schematics in Fig.\,3(e),(f). In addition, the latter site displays a strong magnetic signal, which allows to clearly resolve the two rotational 1$Q$ domains in Fig.\,3(a) with a maximum contrast of about 60\,pm for the right 1$Q$ domain, compared to about 7\,pm magnetic corrugation of the SP-STM data in Fig.\,2(a),(b).
The transition region with a width of about 3\,nm shows a $p(2\times 2)$ magnetic superstructure, which is compatible with a 2$Q$ state. The qualitative agreement  can be seen by direct comparison with a proposed 2$Q$ configuration in Fig.\,3(d), where an orientation of the 2$Q$ state is chosen which is in accordance with the calculated DW in Fig.\,2(f). The atomic spin-resolution achieved here thus supports our analysis based on SP-STM data with a complementary method.

Figure 3(b) shows the central area of Fig.\,3(a), but imaged at a 60$^\circ$ rotated angle, scanning parallel to the AFM rows of the right domain, see arrow in Fig.\,3(b). The similarity of both images means that the measured magnetic signal is largely unaffected by the scan direction and instead is dominated by the surface spin directions with respect to the tip magnetization, as in standard SP-STM.
Careful comparison of Fig.\,3(a) and (b), however, reveals that the lateral positions at which the Co atom jumps to the next hollow site alternate on the right domain in Fig.\,3(a), which is reflected by the different plaquette sizes marked in blue and yellow in Fig.\,3(c), see also Fig.\,S9~\cite{suppl}. The corresponding plaquettes in Fig\,3(b) have almost the same size, i.e.\ the force necessary to move the Co atom is spin-dependent along path\,\textit{A}, and to a lesser degree or not at all along path\,\textit{B}. This means that the corresponding friction becomes anisotropic due to the symmetry of the magnetic texture. We speculate friction to be higher along path\,\textit{B}, where the Co spin is frustrated on all bridge sites and flips between each hollow site, compared to path\,\textit{A} where it flips only every second jump~\cite{wolterPRL2012}. A similar effect can be expected for the 2$Q$ state, but the corresponding differences of Figs.\,3(a) and (b) are too small for a meaningful evaluation, and we propose further experimental investigations in the spirit of Refs.\,\onlinecite{ternesS2008,brandPRB2018}.

\begin{figure}[t] 
	\centering
	\includegraphics[width=0.99\columnwidth]{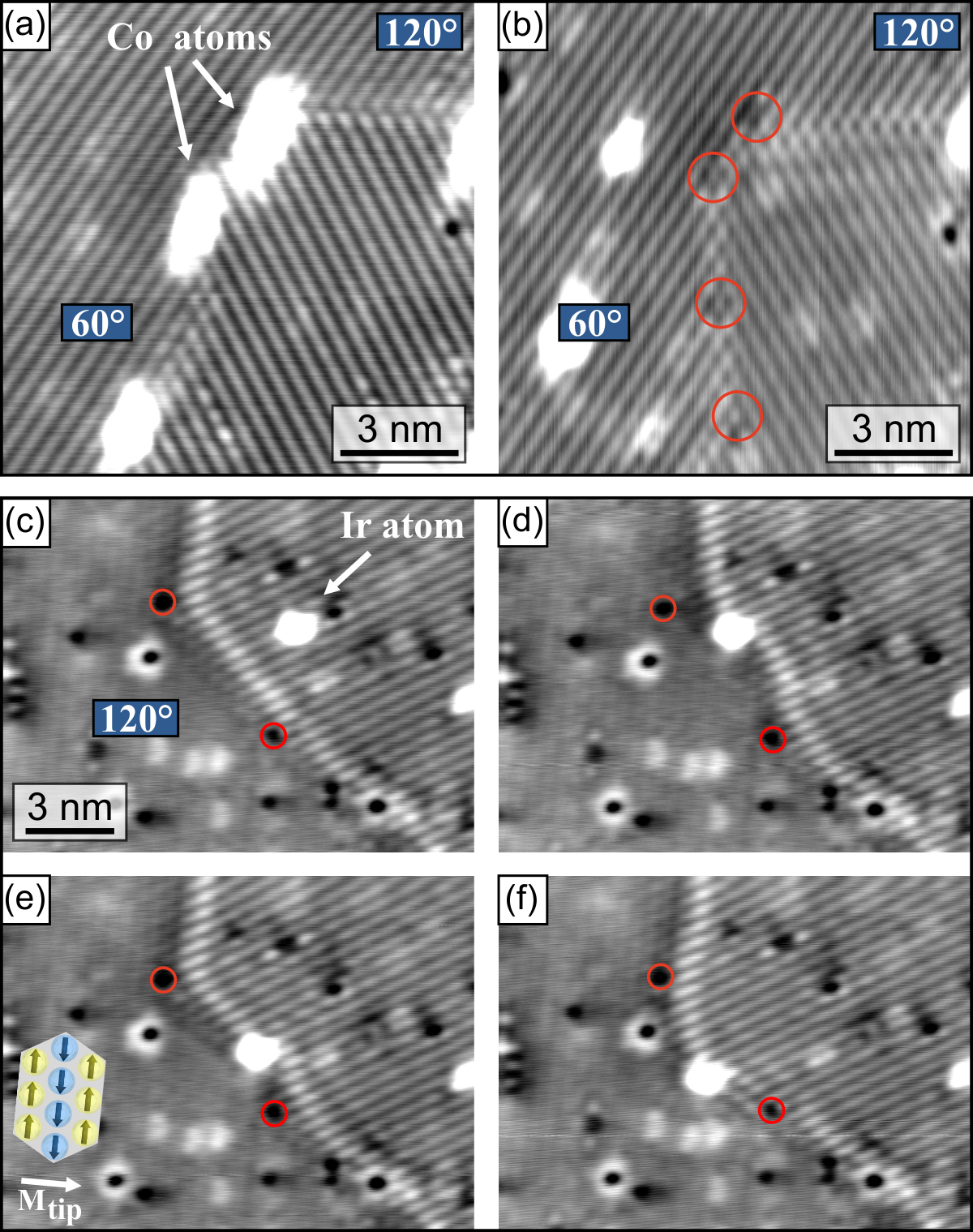}
	\caption{Domain wall control by atom manipulation. (a)~SP-STM image of a DW decorated with Co atoms or clusters (data of Fig.\,2(a), $U=+14.8$\,mV, $I=5$\,nA). (b)~Same area after moving the Co atoms to the left. DW position and tip have changed in the manipulation process. Red circles mark native pinning sites ($U=+19.8$\,mV, $I=5$\,nA, manipul.: $U=+4$\,mV and $I=40$\,nA). (c)~Ir atom and 120$^\circ$\,DW. Magnetic contrast on left domain is very low for this particular tip magnetization direction, see schematics in (e). Two native defects are marked as reference points. (d)-(f) The DW position changes when the Ir atom is moved, demonstrating AFM DW control by atom manipulation; full manipulation series is shown in Fig.\,S10~\cite{suppl} ($U=+10$\,mV, $I=2$\,nA, manipul.: $U=+3$\,mV and $I=60$\,nA).}
		
	\label{fig4}
\end{figure}

While Fig.\,3 shows that the AFM spin texture affects the movement of a Co atom, in Fig.\,4 we demonstrate the inverse case: the position of DWs can be changed by manipulating individual atoms. Figure 4(a) shows again Fig.\,2(a) with Co atoms decorating a DW. When the Co atoms are removed in Fig.\,4(b), the DW relaxes to a slightly different position, pinned to native defects: the 120$^\circ$\,DW section now deviates from the high symmetry direction and the 60$^\circ$\,DW section has moved to the right.
In a second example in Figs.\,4(c)-(f) we use Ir as an adatom, which is non-magnetic as bulk material. For the left domain the magnetic contrast almost vanishes for this particular tip magnetization direction, see schematics in Fig.\,4(e). Two native defects are marked as reference points.
When the Ir atom is positioned closer to the DW in Fig.\,4(d), the DW moves toward the Ir atom. Different DW positions can be achieved with different Ir atom positions in Figs.\,4(d)-(f), overall indicating an attractive interaction between DW and Ir atom, which competes with DW pinning to native defects. When the Ir atom is moved too far away, the DW relaxes to its original position of Fig.\,4(c), see Fig.\,S10~\cite{suppl}. On first glance it is surprising that magnetic DWs can be controlled by non-magnetic atoms. However, most native defects are also non-magnetic atoms like C and O, and their pinning potential is apparent, e.g.\ in Fig.\,4(b); pinning can be achieved by a local change of one of the magnetic interactions \cite{hannekenNJP2016}, and does not necessarily require a magnetic moment.
In any case, the ability to alter, control and prepare specific AFM spin configurations allows a deeper insight into the magnetic properties of AFM systems and might open up new possibilities to investigate the interplay of complex spin textures with the superconducting Re substrate below $T=1.7$\,K~\cite{bedrowARXIV2020,wangS2020}.

In summary, we have demonstrated that AFM domain walls can be imaged with an STM on different length scales, by either measuring density of states, or by employing standard SP-STM or magnetic atom manipulation imaging. The low symmetry of the row-wise AFM spin texture has a number of consequences which might not be restricted to hexagonal surfaces~\cite{ferrianiPRL2007}: (i) the DWs exhibit a distinct 2$Q$ spin texture which should give rise to large signals in transport measurements, (ii) the avoidance of FM edges promotes domain formation, and (iii) the interaction between hexagonal surface and magnetic adatoms becomes anisotropic, which might affect atom diffusion and growth processes.

A.K.\ and K.v.B.\ acknowledge financial support from the Deutsche Forschungsgemeinschaft (DFG, German Research Foundation) Grants No.~408119516 and No.~418425860. R.W.\ acknowledges financial support from the ERC (Adv.\ Grant ADMIRE). A.K.\ thanks J.\ Hagemeister for adding ASE to the simulation code~\cite{github} and M.\ Bazarnik, R.\ Lo Conte, and B.\ Wolter for discussions.

\end{document}